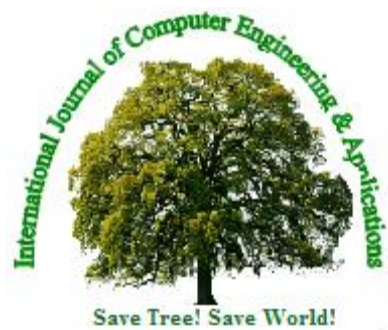

# A SURVEY ON VIDEO FORGERY DETECTION

Sowmya K.N. [1], H.R. Chennamma [2]

[1]Asst. Prof., Department of ISE, JSSATE, Bangalore,  [2]Asst. Prof., Department of MCA , Sri Jayachamarajendra College of Engg., Mysuru, Visvesvaraya Technological University ,Karnataka, India

**ABSTRACT:**

*The Digital Forgeries though not visibly identifiable to human perception it may alter or meddle with underlying natural statistics of digital content. Tampering involves fiddling with video content in order to cause damage or make unauthorized alteration/modification. Tampering detection in video is cumbersome compared to image when considering the properties of the video. Tampering impacts need to be studied and the applied technique/method is used to establish the factual information for legal course in judiciary. In this paper we give an overview of the prior literature and challenges involved in video forgery detection where passive approach is found.*

# [1] INTRODUCTION

The video file generated by the camera as a specified extension depending on the container, which describes the structure of file with the help of metadata. The content of the video is an encoded stream of bytes i.e. codec; which is the chief determiner of quality. Popular container can hold many codec's, for example .MOV file containing H.264 data. Video is basically group of pictures (GOP) organised into frames in a specific sequence. It is a collection of consecutive frames or GOP's with temporal dependency in a three dimensional plane. All frames within a single camera action comprise a shot. A scene is one or more shots that form a semantic unit. Encoding a video involves operation such as quantization which leaves a unique fingerprint on the sequence itself.

Intentional modification/alteration of the digital video for fabrication is referred to as digital video forgery. Implication of it depends upon the circumstance and where it is used. Particularly in the movie, political and medical world its impact is enormous where it is used to defame a personality, hide or forge important information to falsify or conceal actual. Wide usage or sharing of videos in social media like WhatsApp, Youtube, Facebook and news channels has a huge impact in our daily lives. "Seeing is no longer believing" [1]. Integrity and authenticity of video being displayed cannot be accepted blindly anymore.  Growth in video tampering has a multidimensional effect on the stakeholders involved. In order to prove its integrity various forgery detection schemes have been devised by several scientists working in this domain.

Video forgery detection can be broadly classified into active and passive based approaches.

In active approach digital video undergoes some pre processing like watermark insertion, digital signatures etc which would degrade the performance of the video to certain extent at the time of creation of the video. If the video is forged then recovery of the watermark or signature would not be possible and identification of it will result in tampering detection. Active approach cannot be used in scenarios where there is no pre-processing insertions done (watermark or signature absent).

Passive approach works on the basic assumption that video contains naturally occurring properties or inherent fingerprint which is unique to it due to different video imaging devices and its characteristics. If the video is not forged then the underlying statistical correlations of the given video remains consistent after some post processing operations. Identifying the difference, before and after video acquisition which may be malicious or non malicious alteration is the main task of various detection methodologies. In most cases passive forensics can be converted to a problem of pattern recognition [2]. [Figure 1] shows the passive approach for video forgery detection.

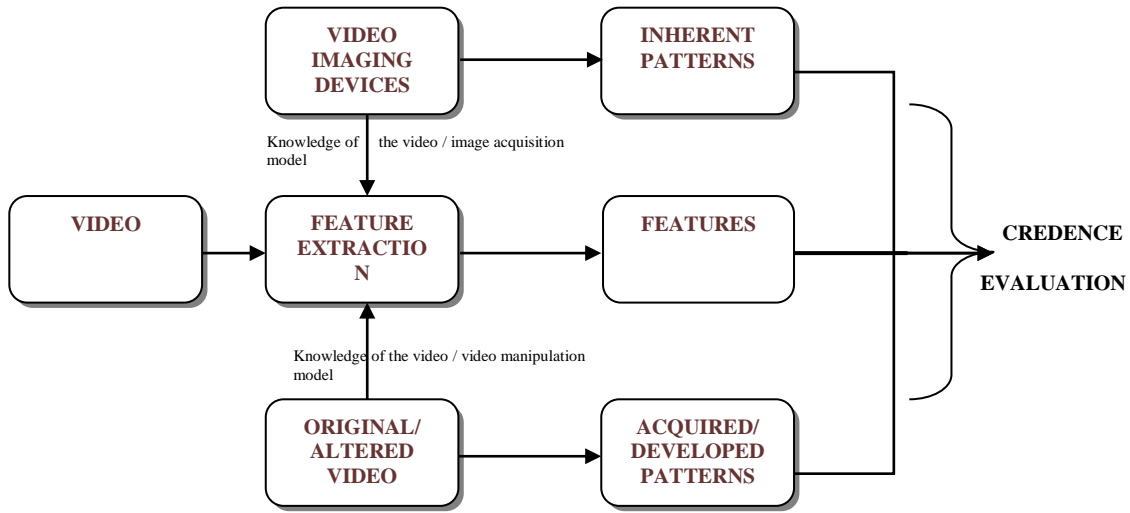

**Figure: 1. Passive approach for video forgery detection**

If a given video integrity need to be checked then several questions need to be answered like, which device (camera, scanner or Computer Graphics) generated the GOPs or video, is it generated as claimed, its processing history and techniques involved [3]. Most of the time the approach we follow is blind in real life since there will not be any recognized authentic video to compare the given. Feature extraction in all cases plays an important role in determining the authenticity. When we say a video is tampered then it contains two parts original part and the altered or tampered part. Researchers have proposed various methods and algorithms to identify digital video tampering based on different features.

Video tampering attacks can be classified based on the boundary of its occurrence into
1. Spatial tampering attacks
2. Temporal tampering attacks and
3. Spatio - temporal tampering attacks

**[Figure 2]** represents video tampering at the broader level.

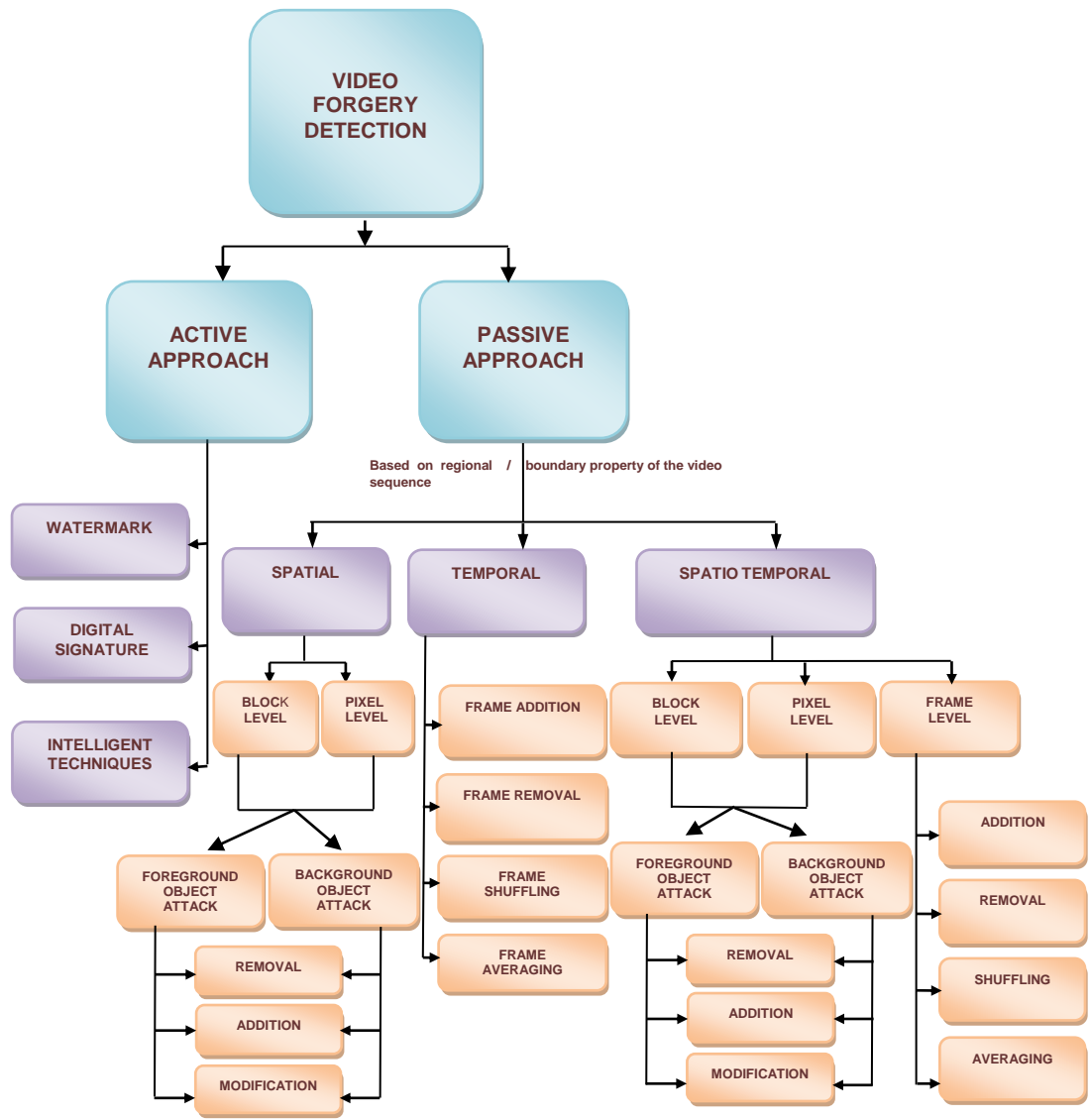

**Figure: 2. Video forgery (tampering) classification**

# [2] SPATIAL TAMPERING

Alterations are performed on content of the frame (x- y axis), which presents visual information of the video. Operations for spatial tampering are cropping & replacement, morphing, inpainting, modifying, content addition and removal. Spatial tampering can be done at block or pixel level. In both cases objects of video frames are altered.

**[2.1] BLOCK LEVEL:**

Tampering attacks are performed on blocks of video frames. Block is basically a specified area on the frame of the video. Blocks can be cropped, replaced, morphed, and modified in any way in block level tampering. Luo et al. [4] proposed a method to detect the changes of GOP based on the blocking artefacts of sequential frames.

**[2.2] PIXEL LEVEL:**

Contents are modified at pixel level. It is the smallest level in video sequences at which tampering attack can be performed. Video authentication system should be robust enough to differentiate the normal video processing operation and pixel level tampering. Normal video processing operations are performed at pixel level. Cloning (copy-move), splicing and re-sampling are the common pixel level operations found.

Spatial tampering detection is also roughly categorized by Farid [5] into:

*Pixel based* which include copy move, splicing, resampling.

Chen et al. [6] have proposed video sequence matching based on temporal ordinal measurement which helps in video copy detection both at spatial and temporal levels.

*Format based* like double mpeg compression, mpeg blocking.

Tanfeng et al. [7] proposed a method based on MPEG double compression. Double compression will disturb Benford's law for $1^{st}$ digit distribution of quantized AC coefficients. Serial SVM is used to estimate the original bit rate scale in doubly compressed video. Milani et al. [8] considered number of compression steps (up to 3) applied to a video as evidence for video editing where classifiers based on multiple SVM were used. Chen et al. [9] modelled generalized Benford's law to detect double compression in MPEG coded video. In the technique proposed the basic idea is that in a recompressed video the statistics of quantized or inverse quantized coefficients exhibit a deviation from that of original video and this difference is used to detect double compression. Disadvantage associated with this method is the edited (forged) raw frames which get single compressed while the rest get double compressed cannot be detected.

*Hardware or camera based tampering* which include CFA, sensor noise, camera response function, chromatic aberrations, demosaicing, white balancing.

Popescu [10] used EM algorithm successfully to identify the demosaicing algorithm used by a particular camera. Mondaini et al.[11] detected video forgery with sensor pattern noise. Johnson et al. [12] exploited camera's optical system by considering lateral chromatic aberrations which gets disturbed when image is forged or tampered. A block that deviated significantly from the global estimate was suspected to be tampered. Drawback is that they have assumed only a small portion of an image is tampered and it will not affect the global estimate. Hyun [13] proposed a technique to detect forgeries based on sensor pattern noise in surveillance videos. Upscale-Crop and partial manipulations forgeries are found with the help of MACE-MRH correlation filters by exploiting the scaling tolerance or invariance. Static scene videos have given impressive results over others. This method gave superior results for RGB/IR videos, static/dynamic videos and compressed videos too.

*Physics based video tampering detection* where various properties like phase congruency, light direction and environment is considered.

Ravi et al. [14] demonstrated the presence of ENF signals in video recordings and lightings using optical sensors. It is used as a natural timestamp for video recordings under fluorescent lighting indoors. Statistical correlations are high with respect to ENF signals from the main supply when video is un tampered and discontinuities are found when it is tampered, which indicates temporal tampering.

*Geometry based like* principal point and metric measurement.

Timothy et al. [15] proposed motion interpolation technique. Modified mean shift mechanism allows a video tube separated into several layers, played in different speeds and then merged. Motion interpolation is proposed based on the reference stick figures and a video inpainting mechanism.

In both block and pixel levels various background and foreground operations can happen.

### [2.3] OBJECT REMOVAL ATTACK:

Here objects of video are eliminated, for example person elimination etc. This kind of attack happen in a specific time domain, recorded in video. This attack can be performed with both foreground and background object. Hsu et al. [16] proposed a method aimed at tampering localization and video splicing was detected where an object is removed from a scene by analyzing the sensor noise characteristics. This method is suitable for detecting or capturing video inpainting tampering and not suitable for detecting other type of forgeries where frame rate is low and edited and on an uncompressed form.

### [2.4] OBJECT ADDITION ATTACK

Here an object of interest can be inserted into a frame or to a set of frames; then it is spatial tampering attack. For example in video evidence, such act with the help of video editing software can mislead court of law. It can be performed with both foreground and background object. Kobayashi et al. [17][18] detected video forgeries using sensor noise patterns from static scene videos. Their proposed method depended on the codec used for video compression because of artefact noise and noise reduction. Splicing attacks are considered where a portion of a video sequence is copied on to another. Detector exploits the differences of noise characteristics between the original and the spliced sequence, being sensitive to compression. Disadvantage is that they can capture only copy-paste related forgery or video inpainting tampering and not suitable for detecting other types of forgeries where frame rate is low and edited on uncompressed form.

### [2.5] OBJECT MODIFICATION ATTACK:

Modification can exist in many prospects in the given video. Size and shape of the existing object may be changed. Colour may be changed or discoloured. With the help of additional effect, nature of object and its relation with other object also may be changed. These attacks are performed generally at pixel level, For example changing face of a person to introduce new face. It can also be performed with both kinds of foreground and background object. Wang et al. [19] exposed intra video forgery based on video content. Regions or frames are replaced with duplicates from the same video sequence to hide unfavourable objects in a scene by overwriting these with background from other segments in the same video. Frame duplication and region duplication are detected by means of analysis of correlations between original regions and cloned ones; Detecting unnatural high coherence is useful for discovering copy paste tampering. Since high coherence is found when region duplication is concerned, pair of tampered frames is assumed to be known a –priori. Global correlations are proposed which takes less time when compared. Drawback is that it cannot be used to detect superimposition caused by inserting objects from other videos.

## [3] TEMPORAL TAMPERING

It is performed on sequence of frames. Focus on temporal dependency. Such attacks are mainly affecting time sequence of visual information, captured by video recording device. Common attacks are: frame addition, Frame removal, Frame reordering, Frame shuffling.

Temporal tampering can be at scene level, shot level, and frame level. Primary focus is on attacking the temporal dependency of frames of video called as `third dimensional attack' with respect to time on the video sequence. Wang et al. [20] proposed a technique by analyzing the periodic properties of predictive 'P' frame prediction errors in MPEG-2 video to detect double compression. It can only detect double compressed intra frame 'I' in variable bit rate mode (i.e. Constant quantization scale factor). Wang et al. [21] also proposed a technique to detect double compression by capturing empty bins exhibited in the distribution of quantized coefficients in a recompressed video. Drawback of this approach is that it cannot detect if one or more bidirectional interpolated predictive frame 'B' or P frames are authentic or forged when the frame rate is low and the existence of evidence is captured in only a few frames. When the forged frames are edited in uncompressed form and then compressed they get single compressed while the rest double compressed, such frames cannot be detected. Shan et al. [22] proposed an efficient feature vector to expose fake high bit rate videos. They tried to estimate the original bit rate by analyzing the re quantization artefacts in 'I' frames as well as the quality change between a given video and its sequential bit rate down converted versions. These features are then combined with other video properties such as adaptive rate control schemes in video coding. Drawback of this method is that fake bit rate videos tested are recompressed with the same codec. Exposing those resulting video with different codec is the present requirement to work on.

## [3.1] FRAME ADDITION ATTACK:

Additional frames from another video, which has same statistical property, are intentionally inserted at some random locations in a given video. It intends to camouflage the actual content and provide incorrect information.

Gironia et al. [23] proposed inter frame forgery detection for various codec based videos where frames are inserted or deleted as part of tampering. Double encoding detection is done using VPF tool and sliding window analysis (moving) helps to identify the phase discontinuity of a periodic function if number of frames is removed. Localization of frames insertion helps to detect 2 cutting points, one at the beginning and the other one at the end of inserted frame sequence. Their proposed method cannot detect frame manipulation when whole GOP is removed /inserted.

### [3.2] FRAME REMOVAL ATTACK

Frames are intentionally eliminated. In this kind of attack, frames can be removed from a specific location or it can be removed from different locations. Depending upon the intention it is generally performed on surveillance video where the intruder wants to remove his/her presence.

Chao et al. [24] proposed to detect frame deletion and insertion through optical flow. Inter frame forgery would cause discontinuity in optical flow sequence. Yuxing et al. [25] proposed Video inter frame forgery detection based on discontinuities found in the relative factor sequence (VFI) after maximum sampling based on generalized ESD test which follows approximate normal distribution. Sequence varies for deletion and duplication.

### [3.3] FRAME SHUFFLING ATTACK

Here frames of a given video are shuffled or reordered in such a way that correct frame sequence is intermingled and wrong information is produced by the video when compared to original video. Yuting et al. [26], Stamm et al. [27] have proposed inter frame forgery detection by exploiting the frame grouping strategy adopted by all common video encoders.

### [4] SPATIO TEMPORAL TAMPERING

It is a combination of temporal and spatial tampering. Frame sequences and visual contents are modified in the same video. It is done at scene level. Intra frame tampering and inter frame tampering combination is found here. Various tampering methods found in spatial and temporal tampering domains are also found here.

Bestagini et al. [28], have proposed algorithm/method that is able to detect whether a spatio temporal region of a sequence or GOP was replaced by series of fixed repeated images in time by analyzing the footprints left on the residual computed between adjacent frames which is robust to mild compression, or a portion of the same video is taken at different interval of time by exploiting correlation analysis. Tampering localization fails when the duplicated block is not correctly identified. Subramanyam et al. [29] detected splicing from the same frame by matching histogram of oriented gradients (HOG). Compositing where an object from one frame is copied to another frame is detected by exploiting MPEG-2 GOP structure. Connotter et al. [30] exposed forgery based on video content. Their proposed technique explicitly models the three-dimensional ballistic motion of objects in free-flight and the two-dimensional projection of the trajectory into the image plane of a static or moving camera. Deviations from this model provide evidence of manipulation. Subramanyam et al. [31] proposed an algorithm which uses principal of estimation theory to detect double quantization. CRLB is applied or attained to find the efficient estimator for the likelihood function. It can detect tampering of 'I', 'P' or 'B' frames with high accuracy. Proposed technique can detect forgery under wide range of double compression bit rates or quantization scale factors.

Wang et al. [32] uncovered digital forgeries in interlaced and deinterlaced video by detecting spatial and temporal localized tampering. Interlaced video creates spatial artifacts for moving object at fast rate. Their model measures, interfield and interframe motion by performing motion estimation and spatial/temporal differentiation. Motion estimation remains same for both, if original video otherwise tampered. Deinterlaced video correlations are explicitly modeled by field extension algorithms like that of line repetition, line averaging, field insertion & vertical temporal interpolation algorithm, motion adaptive and motion compensated estimation algorithm. Once the deinterlaced videos are modeled using any of these correlation algorithms spatial and temporal correlations are found by choosing linear model and EM algorithm is used. Tampering can be locally identified both in time and space. Both algorithms are used to detect frame rate conversion which varies if a video is altered or manipulated. Limitations are the correlation estimation success is based on compression artifacts for an interlaced and a deinterlaced video. A deinterlaced video may be tampered initially and then interlaced for which a de-interlacing algorithm may be applied such that correlation appear to remain intact.

Object based video forgery detection is done by Richao et al. [33] where statistical property of an object and variable width of its neighbor area is considered with adjustable width object boundary (AWOB) defined. Non sub sampled contourlet coefficients and gradient information exploits the image edge locally with the help of Gaussian distribution model. Edge intensities are analyzed with the help of Rayleigh distribution. SVM classifier (RBF of libsvm library) is used to separate the natural objects and forged ones. Main drawback is that the trained sample database is required. Standard Video forensic dataset too is still in need. Renjie et al. [34] addressed an efficient spatio temporal scheme to extract moving objects from video sequence. During temporal segmentation moving objects are localized and examined with the help of block based motion detector to identify the changed region. Slow and fast motion of objects is considered. Once temporal segmentation is complete, identify the static and dynamic regions, spatial segmentation is done with the help of watershed algorithm. A combination of these segmentations with a fixed threshold is then used to detect desired moving objects. Results are not much impressive where the video sequence is short. Identifying the objects with moving background using global motion estimation needs to be addressed further.

Thus Digital video tampering can be done at various stages; Scene level, shot level, frame level, block level and pixel level or a combination of these levels. **[Figure 3]** represents video tampering relationship. Depending on the tampering level we classify them into temporal, spatial and spatio temporal attacks. Currently maximum work is found in spatial tampering detection since it is close to image forgery detection.

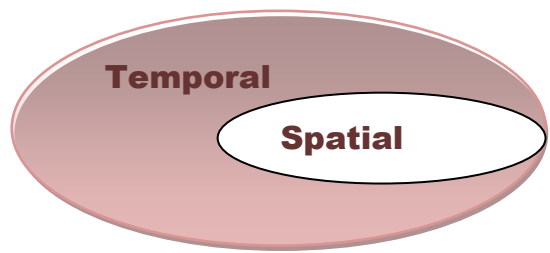

**Figure: 3. Spatio temporal relationship**

Passive tampering detection schemes include learning based schemes, threshold based schemes and schemes based on compression properties. Coding bit rate has a significant impact on the efficiency of the algorithm since lower the quality of video, longer the sequence to be analyzed. We need to understand how a specific form of tampering disturbs certain statistical properties of the video and then develop a mathematical model /algorithm to detect this perturbation.

## [5] CONCLUSION

Digital video forensics is still in its infant stage. Reliability of the digital video as a reference to reality is under threat due to tampering. Various editing tools like Adobe's Photoshop & illustrator, GNU Gimp, Premier, Vegas are available easily [35]. Research community needs to establish various methods or tampering detection schemes at all levels of video classification. Of the many kinds of perturbations that tampering can introduce into a video, we have explored and listed few works till recent. With respect to passive technology, selecting the key patterns are deterministic in determining whether the video is tampered or not. The more developed field of image forensics can provide a source of new ideas for future video forensic tools. Standard video dataset containing forged and authenticated videos under various conditions are the requirement of the day for researchers to evaluate their proposed methods. We need to combine together various proposed approaches to detect many kinds of tampering operations.

Improvement in the field of digital video forensics particularly tampering detection will help us in improving the reliability factor for digital video and we will be able to understand their properties in a much better way in the field of entertainment industry, surveillance video, medical field and legal concerns. New methods combining both active and passive based approach will rule the future video forensics era with the domain knowledge or expertise involving pattern recognition, artificial intelligence, signal processing, and machine learning & computer vision [36].

REFERENCES

[1] Zhub, Swanson M, Tewfik A. when seeing isn't Believing [multimedia authentication technologies.], IEEE signal processing magazine , Vol 21, No(2), pp40-49, Mar 2004

[2] LUO Weiqi,QU Zhenhua,PAN Feng, HUANG Jiwu. A survey of passive technology for digital image forensics [J]. Front. Comput. Sci., 2007, 1(2): 166-179

[3] Sencar, H. T., & Memon, N. Overview of state-of-the-art in digital image forensics. *Algorithms, Architectures and Information Systems Security*, *3*, 325-348, 2008.

[4] Luo, Weiqi, Min Wu, and Jiwu Huang. "MPEG recompression detection based on block artifacts." *Electronic Imaging*. International Society for Optics and Photonics, 2008.

[5] H.Farid Farid, Hany. "Image forgery detection." *Signal Processing Magazine, IEEE* 26, no. 2 (2009): 16-25.

[6] Chen, Li, and F. W. M. Stentiford. "Video sequence matching based on temporal ordinal measurement." *Pattern Recognition Letters* 29, no. 13 (2008): 1824-1831.

[7] Sun, Tanfeng, Wan Wang, and Xinghao Jiang. "Exposing video forgeries by detecting MPEG double compression." In *IEEE International Conference on* Acoustics*, Speech and Signal Processing (ICASSP), 2012*: pp. 1389-1392.

[8] Milani, Simone, Paolo Bestagini, Marco Tagliasacchi, and Stefano Tubaro. "Multiple compression detection for video sequences." In IEEE *14th International Workshop on Multimedia Signal Processing (MMSP), 2012:* pp. 112-117.

[9] Chen, Wen, and Yun Q. Shi. "Detection of double MPEG compression based on first digit statistics." In *Digital Watermarking*, pp. 16-30. Springer Berlin Heidelberg, 2009.

[10] Popescu, Statistical tools for digital image forensics. PhD thesis, *Dept of Computer. Sci., Dartmouth College*, Hanover, USA, 2004.

[11] Mondaini, N., Roberto Caldelli, Alessandro Piva, Mauro Barni, and Vito Cappellini. "Detection of malevolent changes in digital video for forensic applications." In *Electronic Imaging 2007*, pp. 65050T-65050T. International Society for Optics and Photonics, 2007.

[12] Johnson, Micah K., and Hany Farid. "Exposing digital forgeries through chromatic aberration." In *Proceedings of the 8th workshop on Multimedia and security*, pp. 48-55. ACM, 2006.

[13] Hyun, Dai-Kyung, Min-Jeong Lee, Seung-Jin Ryu, Hae-Yeoun Lee, and Heung-Kyu Lee. "Forgery detection for surveillance video." In *The Era of Interactive Media*, pp. 25-36. Springer New York, 2013.

[14] Garg, Ravi, Avinash L. Varna, and Min Wu. "Seeing ENF: natural time stamp for digital video via optical sensing and signal processing." In *Proceedings of the 19th ACM international conference on Multimedia*, pp. 23-32. ACM, 2011.

[15] Shih, Timothy K., Nick C. Tang, Joseph C. Tsai, and Jenq-Neng Hwang. "Video motion interpolation for special effect applications." *Systems, Man, and Cybernetics, Part C: Applications and Reviews, IEEE Transactions on* 41, no. 5 (2011): 720-732.

[16] Hsu, Chih-Chung, Tzu-Yi Hung, Chia-Wen Lin, and Chiou-Ting Hsu. "Video forgery detection using correlation of noise residue." In IEEE *10th Workshop on Multimedia Signal Processing, 2008*, pp. 170-174.

[17] Kobayashi, Michihiro, Takahiro Okabe, and Yoichi Sato. "Detecting video forgeries based on noise characteristics." In *Advances in Image and Video Technology*, pp. 306-317. Springer Berlin Heidelberg, 2009.

[18] Kobayashi, Michihiro, Takahiro Okabe, and Yoichi Sato. "Detecting forgery from static-scene video based on inconsistency in noise level functions." *Information Forensics and Security, IEEE Transactions on* 5.4 (2010): 883-892.

[19] Wang, Weihong, and Hany Farid. "Exposing digital forgeries in video by detecting duplication." *Proceedings of the 9th workshop on Multimedia & security*. ACM, 2007.

[20] Wang, Weihong, and Hany Farid. "Exposing digital forgeries in video by detecting double MPEG compression." In *Proceedings of the 8th workshop on Multimedia and security*, pp. 37-47. ACM, 2006.

[21] Wang, Weihong, and Hany Farid. "Exposing digital forgeries in video by detecting double quantization." In *Proceedings of the 11th ACM workshop on Multimedia and security*, pp. 39-48. ACM, 2009.

[22] Bian, Shan, Weiqi Luo, and Jiwu Huang. "Exposing Fake Bit-rate Videos and Estimating Original Bit-rates.", 2014.

[23] Gironi, A., M. Fontani, Tiziano Bianchi, A. Piva, and M. Barni. "A video forensic technique for detecting frame deletion and insertion." In *IEEE International Conference on* Acoustics*, Speech and Signal Processing (ICASSP), 2014:* pp. 6226-6230.

[24] Chao, Juan, Xinghao Jiang, and Tanfeng Sun. "A novel video inter-frame forgery model detection scheme based on optical flow consistency." *Digital Forensics and Watermaking*. Springer Berlin Heidelberg, 2013. 267-281.

[25] Wu, Yuxing, Xinghao Jiang, Tanfeng Sun, and Wan Wang. "Exposing video inter-frame forgery based on velocity field consistency." In IEEE *International Conference on Acoustics, Speech and Signal Processing (ICASSP), 2014*: pp. 2674-2678.

[26] Zhang, Jing, Yuting Su, and Mingyu Zhang. "Exposing digital video forgery by ghost shadow artifact." In *Proceedings of the First ACM workshop on Multimedia in forensics*, pp. 49-54. ACM, 2009.

[27] Stamm, Matthew C., W. Sabrina Lin, and KJ Ray Liu. "Temporal forensics and anti-forensics for motion compensated video." *Information Forensics and Security, IEEE Transactions on* 7.4 (2012): 1315-1329.

[28] Bestagini, Paolo, Simone Milani, Marco Tagliasacchi, and Stefano Tubaro. "Local tampering detection in video sequences." In IEEE *15th International Workshop on Multimedia Signal Processing (MMSP), 2013:* pp. 488-493.

[29] Subramanyam, A. V., and Sabu Emmanuel. "Video forgery detection using HOG features and compression properties." *In IEEE 14th International Workshop on Multimedia Signal Processing (MMSP), 2012*.

[30] Conotter, Valentina, James F. O'Brien, and Hany Farid. "Exposing digital forgeries in ballistic motion." *Information Forensics and Security, IEEE Transactions on* 7, no. 1 (2012): 283-296.

[31] Subramanyam, A. V., and Sabu Emmanuel. "Pixel estimation based video forgery detection." In IEEE *International Conference on Acoustics, Speech and Signal Processing (ICASSP), 2013:* pp. 3038-3042.

[32] Wang, Weihong, and Hany Farid. "Exposing digital forgeries in interlaced and deinterlaced video." *Information Forensics and Security, IEEE Transactions on* 2, no. 3 (2007): 438-449.

[33] Chen, Richao, Qiong Dong, Heng Ren, and Iiaqi Fu. "Video Forgery Detection Based on Non-Subsampled Contourlet Transform and Gradient Information." *Information Technology Journal* 11, no. 10 (2012).

[34] Li, Renjie, Songyu Yu, and Xiaokang Yang. "Efficient spatio-temporal segmentation for extracting moving objects in video sequences." *Consumer Electronics, IEEE Transactions on* 53, no. 3 (2007): 1161-1167.

[35] Rocha, Anderson, Walter Scheirer, Terrance Boult, and Siome Goldenstein. "Vision of the unseen: Current trends and challenges in digital image and video forensics." *ACM Computing Surveys (CSUR)* 43, no. 4 (2011): 26.

[36] Ng T T, Chang S F, Lin C Y, et al. Multimedia Security Technologies for Digital Rights, chap. Passive-Blind image forensic, Elsevier, 2006.